\documentclass{article}
 \usepackage{amsmath}
 \usepackage{graphicx}
 \usepackage{amsthm}
\usepackage{hyperref}
\usepackage{algpseudocode}
\usepackage{algorithm}
 \usepackage{epstopdf}
 \usepackage{subfigure}
 \usepackage[font=small]{caption}
  \usepackage{authblk}
   \usepackage[T1]{fontenc}
  \usepackage[utf8]{inputenc}
  \usepackage{fix-cm}

   \usepackage{enumerate}
   
\date{\vspace{-5ex}}

\makeatother
 
\begin{document}
 \title{Sparse Estimation with Generalized Beta Mixture and the Horseshoe Prior}
 \author{Zahra Sabetsarvestani \footnote{Amirkabir University of Technology. \url{zahra.sabet@aut.ac.ir}}, Hamidreza Amindavar\footnote{Amirkabir University of Technology. \url{hamidami@aut.ac.ir}}}

 \maketitle
 \begin{abstract}
In this paper, the use of the Generalized Beta Mixture (GBM) and Horseshoe distributions as priors in the Bayesian Compressive Sensing framework is proposed. The distributions are considered in a two-layer hierarchical model, making the corresponding inference problem amenable to Expectation Maximization (EM). We present an explicit, algebraic EM-update rule for the models, yielding two fast and experimentally validated algorithms for signal recovery. Experimental results show that our algorithms outperform state-of-the-art methods on a wide range of sparsity levels and amplitudes in terms of reconstruction accuracy, convergence rate and sparsity. The largest improvement can be observed for sparse signals with high amplitudes.
 \end{abstract}
 \section{Introduction}
   Compressive Sensing is a new signal processing framework  for efficiently acquiring and reconstructing a signal that have a sparse representation in a fixed linear basis. Consider the following acquisition system:
   \begin{equation}
   \mathbf{y}=\Phi \mathbf{w} + \mathbf{n}
    \label{(2)}
    \end{equation}
 where $ \mathbf{y} $ is the $ M\times 1 $ measurement vector, $\Phi $  is the $ M\times N $ measurement matrix where $ M < N $ and $ \mathbf{n} $  represents the measurement noise. In this expression $ \mathbf{w} $ is the unknown signal which is considered to be sparse.
    By exploiting the sparsity constraint on $ \mathbf{w} $ the inversion of \eqref{(2)} which is an ill-posed problem is the solution of the following optimization problem:
 \begin{equation}
 \hat{\mathbf{w}}=\textnormal{arg}\min_{\mathbf{w}}{\Arrowvert \mathbf{y}-\Phi \mathbf{w} \Arrowvert_2}+\tau\Arrowvert \mathbf{w} \Arrowvert_0 \label{(3)}
 \end{equation}
 where $ \Arrowvert \mathbf{w} \Arrowvert_0 $ is equal to the number of non-zero entries of $ \mathbf{w} $. This optimization problem is NP-Hard and can not be solved in polynomial time, but it has been shown in~\cite{1} that under some certain conditions on the measurement matrix $ \Phi $, \eqref{(3)} can be replaced by the following convex optimization problem.
 \begin{equation}
 \hat{\mathbf{w}}=\textnormal{arg}\min_\mathbf{w}{\Arrowvert \mathbf{y}-\Phi \mathbf{w} \Arrowvert_2 + \tau \Arrowvert \mathbf{w} \Arrowvert_1} \label{(4)}
 \end{equation}
 Another class of CS recovery methods uses a greedy algorithm to estimate the sparse signal. In the greedy algorithms the support of the sparse signal is estimated sequentially and a local optimization in each step estimates the signal coefficients~\cite{2}.\newline The main focus of this paper is on the Bayesian approaches in compressive sensing, known as the Bayesian Compressive Sensing (BCS). BCS which is based on the \textit{Relevance Vector Machine} ({RVM) or \textit{Sparse Bayesian Learning} (SBL) theory~\cite{4,5} was introduced in~\cite{3} for the first time. In BCS, all of the problem components are modeled in the Bayesian framework and a proper inference method is utilized to reconstruct the sparse signal. Developing a sparsity-inducing prior for modeling the sparse signal efficiently is a main research topic in BCS. The priors should be concentrated on zero to model the sparsity and at the same time, have heavy tail to avoid over-shrinkage of larger signal coefficients. The recent work in this area was led to the introduction of new priors~\cite{12,6}. In~\cite{6} a spike and slab prior is considered on the cluster sparse signal coefficients and a Markov Chain Monte Carlo (MCMC) sampling method is utilized for the inference. Since in MCMC a large set of samples are required for the posterior distribution estimation, this method suffers from high computational complexity and low convergence rate in consequence. Another class of sparsity-inducing priors which is widely used in literature~\cite{7,3} exploit an equivalent conjugate hierarchy model. In the hierarchical model a complex prior is decomposed to two or more simple distributions which facilitate the inference procedure. In RVM~\cite{4} and in its adaption to BCS~\cite{3} a student's t-distribution is assigned as a two layer hierarchical prior
 on the sparse coefficients. In the later work~\cite{7} a hierarchical model of Laplace distribution is used. Laplace and student's t-distribution is concentrated near origin which shrink small coefficients toward zero, but due to having light tails, they over-shrink coefficients not close to zero and fail in modeling signals with large amplitude. large-amplitude coefficients arises in applications where signal shows impulsive behaviour~\cite{14}.\newline
 In this paper we suggest to use \textit{ Generalized Beta Mixture} (GBM) and \textit{Horseshoe} distributions as priors in the BCS framework. The GBM has two free parameters which promote its capability in modeling large-amplitude coefficients and handling different levels of sparsity~\cite{9}. For the special choice of the free parameters, GBM results in the Horseshoe distribution~\cite{10,11}. The Horseshoe distribution, with a pole at the origin and having heavy tail lead to a sort of nearly optimal choice as a prior for sparse signals. In this work a two-layer hierarchical model is used for representing the GBM and Horseshoe priors. For signal recovery we suggest an iterative algorithm based on the Expectation Maximization method and two fast greedy algorithms inspired by the algorithm in~\cite{13}. Our algorithms while enjoy the inherent Bayesian framework advantages, by modeling non-zero coefficients more efficiently and imposing sparsity more heavily, result in lower reconstruction error and faster convergence rate in comparison to the other existing algorithms in this framework. In summary our main contribution are:
 \begin{itemize}
 \item Proposing GBM and Horseshoe as priors in the Bayesian compressive sensing framework.
 \item Providing an explicit, algebraic EM-update for two-layer model of GBM and Horseshoe, yielding two efficient algorithms.
 \item Performing an extensive experimental comparison with state-of-the-art methods, demonstrating superiority on a wide parameter range.
 \end{itemize}
The rest of this paper is as follows: in section \ref{section II}, we study the Bayesian framework for compressive sensing problem, detail hierarchical models and analyze their sparsity-inducing properties. In section~\ref{section III} we detail Bayesian inference procedure and propose two fast algorithms. We present simulation results and comparison with the other existing algorithms in section \ref{section IV} and conclude the paper in section \ref{V}.
\section{Bayesian Framework}\label{section II}
 In the Bayesian framework, a probability density function $p(\mathbf{w}| \boldsymbol{\gamma})$ is considered on the sparse signal, which reflects our prior knowledge about the nature of the original signal. In this work we consider $p(\mathbf{w}| \boldsymbol{\gamma})$ to be GBM or Horseshoe distribution. These priors features appealing properties compared to Laplace and student's t-distribution. GBM and Horseshoe are much heavier-tailed, modeling large signals efficiently while shrinking noise-like, small signals toward zero.\newline The measurement vector is also a statistical process where its distribution depends on the distribution of measurement noise. Since we consider the measurement noise as an independent additive zero mean white Gaussian noise with variance equal to $ \beta^{-1} $, therefore the conditional pdf $ p(\mathbf{y}|\mathbf{w},\beta) $ is a multiple of normal distributions.
 \begin{equation}
 p(\mathbf{y}|\mathbf{w},\beta)=\mathcal{N}(\mathbf{y}|\Phi \mathbf{ w}, \beta^{-1})
 \end{equation}

 In the inference procedure for BCS recovery, we need to obtain the joint pdf of all of the problem components $ p(\mathbf{y},\mathbf{w},\boldsymbol{\gamma},\beta)$  which can be decomposed as follows:
 \begin{equation}
 p(\mathbf{y},\mathbf{w}, \beta,\boldsymbol {\gamma})=p(\mathbf{y}|\mathbf{w}, \beta) p(\mathbf{w}|\boldsymbol {\gamma}) p(\boldsymbol{\gamma}) p(\beta)
 \end{equation}
 In this model $ \beta $ and $ \boldsymbol{\gamma} $ are called \textit{hyper-parameters} and the assigned corresponding distributions $ p(\gamma) $ and $ p(\beta) $ are called \textit{hyper-priors}.
 \subsection{Bayesian Hierarchical Model}
     The optimization problem in \eqref{(4)} corresponds to the \textit{Maximum a Posteriori} (MAP) estimation of $ \mathbf{w} $ while  Laplace is considered as a prior distribution on the sparse coefficients.
     \begin{equation}
     P(\mathbf{w}|\lambda)=\frac{\lambda}{2}\  \textnormal{exp}(-\frac{\lambda}{2}\Arrowvert\mathbf{w}\Arrowvert_1)
 \end{equation}
  But since in this formulation, Laplace is not the conjugate prior to the conditional pdf $ p(\mathbf{y}|\mathbf{w},\beta) $ a tractable Bayesian inference may not be performed. To address this problem, Bayesian hierarchical models were proposed~\cite{4}. In a hierarchical model, the joint pdf $ p(\mathbf{w}| \boldsymbol{\gamma})p(\boldsymbol{\gamma}) $ determines the whole sparsity-inducing properties of the Bayesian model. The conditional prior pdf $ p(\mathbf{w}| \boldsymbol{\gamma})$ is a product of Gaussian pdfs: $  p(\mathbf{w}|\boldsymbol{\gamma})=\prod_{i=1}^{N} \mathcal{N}(w_i|0,\gamma_i) $ and the hyper-prior $ p(\boldsymbol{\gamma}) $ is chosen to be a suitable distribution. For example in the second stage of hierarchical student's t-distribution~\cite{3,4} a Gamma hyper-prior is assigned to the precision variable while in hierarchical model of Laplace a Gamma distribution is considered on the variance of normal~\cite{7}.\newline
   
   \subsection{Generalized Beta Mixture and Horseshoe} 
   In this section we represent the hierarchical GBM and Horseshoe distributions in the conjugate manner. When \textit{Beta2} or \textit{Beta prime} distribution is coupled with normal distribution as a prior for location, it results in a Generalized Beta Mixture. 
   \begin{equation}
   p(w_i|\gamma_i)=\mathcal{N}(w_i|0,\gamma_i)\ \ \ \ p(\gamma_i|a,b)= \mathcal{B}(\gamma_i|a,b)
   \end{equation}
 Where $ \mathcal{B} $ denotes the Beta prime distribution and is defined as
 \begin{equation*}
 \mathcal{B}(\xi|a,b)=\frac{\Gamma(a+b)}{\Gamma(a)\Gamma(b)} \frac {\xi^{(a-1)}}{{(\xi+1)}^{(a+b)}}\ \  \xi>0
 \end{equation*}
 and $ \Gamma(.) $ shows the Gamma function. The smaller value of $ a $ causes sharper peak at zero and consequently concentrates more on mass near zero, while the smaller value of  $ b $ yields to a distribution which is heavier-tailed~\cite{9}. 
 Beta prime distribution can be expressed as a scaled combination of two Gamma distributions in a two layer hierarchical structure which leads to a three layer model for GBM as follows~\cite{9}:
  \begin{equation*}
    p(w_i|\gamma_i)=\mathcal{N}(w_i|0,\gamma_i)\ \ \ \
  p(\gamma_i|a,\lambda)=\mathcal{G}(a,\lambda)\ \ \  p(\lambda|b,1)=\mathcal{G}(b,1)
  \end{equation*}
where $ \mathcal{G}(.) $ represents the Gamma distribution.Figure~\ref{H} shows different hierarchical models of GBM.\newline
 \begin{figure}[t]
     \begin{picture}(400,90)
    
    \thicklines
     \put(20,90){\circle{16}}
     \put(20,50){\circle{16}}
     \put(55,70){\circle{16}}
     \put(95,98){\circle{16}} \put(95,42){\circle{16}}
     \put(130,97){\circle{16}}
     \put(130,41){\circle{16}}
     \put(26,85){\vector(2,-1){22}}
     \put(26,54){\vector(2,1){22}}
     \put(61,75){\vector(3,2){27}}
     \put(61,65){\vector(3,-2){28}}
    \put(102,98){\vector(2,0){19}}
    \put(102,42){\vector(2,0){19}} 
    \put(17,87){$ a $}
    \put(17,47){$ b $}
    \put(52,67){$ \gamma $}
   \put(37,45){$ \mathcal{B}(\gamma|a,b)$}
   \put(91,95){$ \gamma_1 $}
   \put(89,39){$ \gamma_N $}
   \put(125,97){$ w_1 $}
   \put(123,41){$w_N $}
   \put(115,112){$ \mathcal{N}(w_1|0,\gamma_1) $}
   \put(115,23){$ \mathcal{N}(w_N|0,\gamma_N) $}
     \put(45,15){Model \it{I}}

         \put(190,70){\circle{16}}
       \put(230,70){\circle{16}}
      \put(280,70){\circle{16}}
      \put(320,98){\circle{16}} \put(320,42){\circle{16}}
      \put(355,98){\circle{16}}
      \put(355,42){\circle{16}}
       \put(280,98){\circle{16}}
     \put(238,70){\vector(2,0){32}}
       \put(198,70){\vector(2,0){22}}
     \put(286,75){\vector(3,2){28}}
     \put(286,66){\vector(3,-2){28}}
     \put(328,98){\vector(2,0){18}}
     \put(328,42){\vector(2,0){18}}
     \put(280,90){\vector(0,-1){12}}
      \put(187,67){$ b $}
      \put(227,67){$ \lambda $}
     \put(277,67){$ \gamma $}
    \put(316,98){$ \gamma_1 $} \put(315,42){$ \gamma_N $}
    \put(350,98){$ w_1 $}
    \put(348,42){$ w_N $}
    \put(277,95){$ a $}
    \put(208,52){$\mathcal{G}(\lambda|b,1)  $}
     \put(255,52){$\mathcal{G}(\gamma|a,\lambda)  $}
     \put(340,112){$ \mathcal{N}(w_1|0,\gamma_1) $}
     \put(340,23){$ \mathcal{N}(w_N|0,\gamma_N) $}
     \put(245,15){Model \it{II}}
     \multiput(95,97)(0,-3){13}{\line(0,1){1}}
       \multiput(130,87)(0,-3){13}{\line(0,1){1}}
    \multiput(95,87)(0,-3){13}{\line(0,1){1}}
    \multiput(320,87)(0,-3){13}{\line(0,1){1}}
    \multiput(355,87)(0,-3){13}{\line(0,1){1}}

       \end{picture}
        \caption {Hierarchical Models of GBM. Model \textit{I} shows the two-layer model and Model \textit{II} shows the three-layer model. }
        \label{H}
       \end{figure}
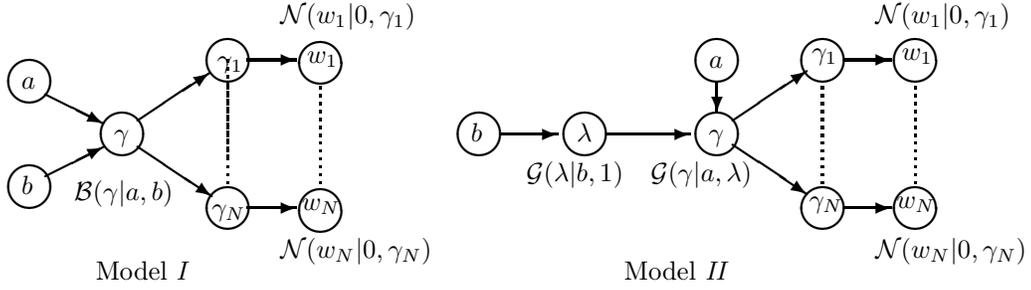
 \begin{figure}[!tb]
  \centering
  
\subfigure[]{\includegraphics[width=45mm,height=30mm]{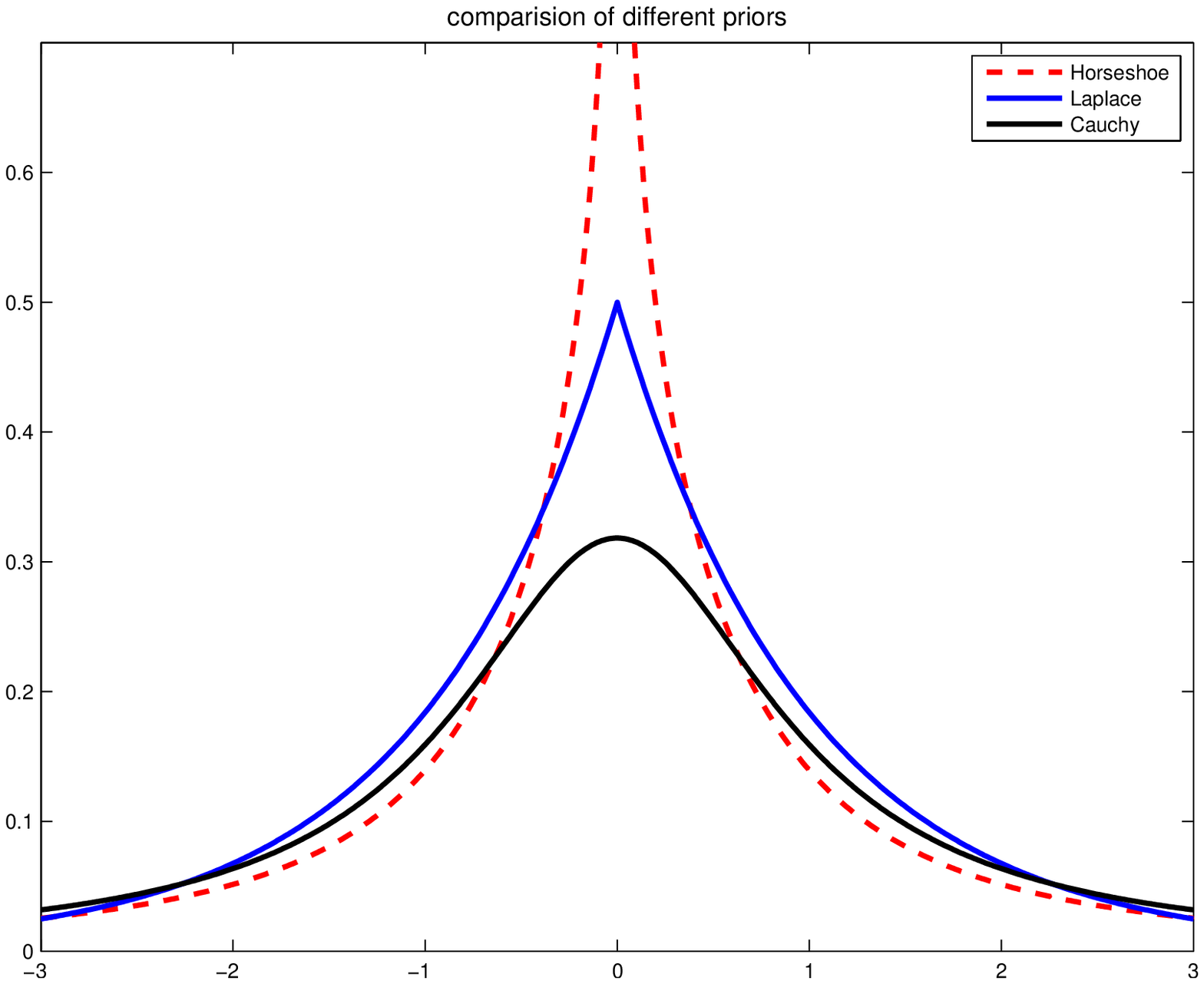}}
\subfigure[]{\includegraphics[width=45mm,height=30mm]{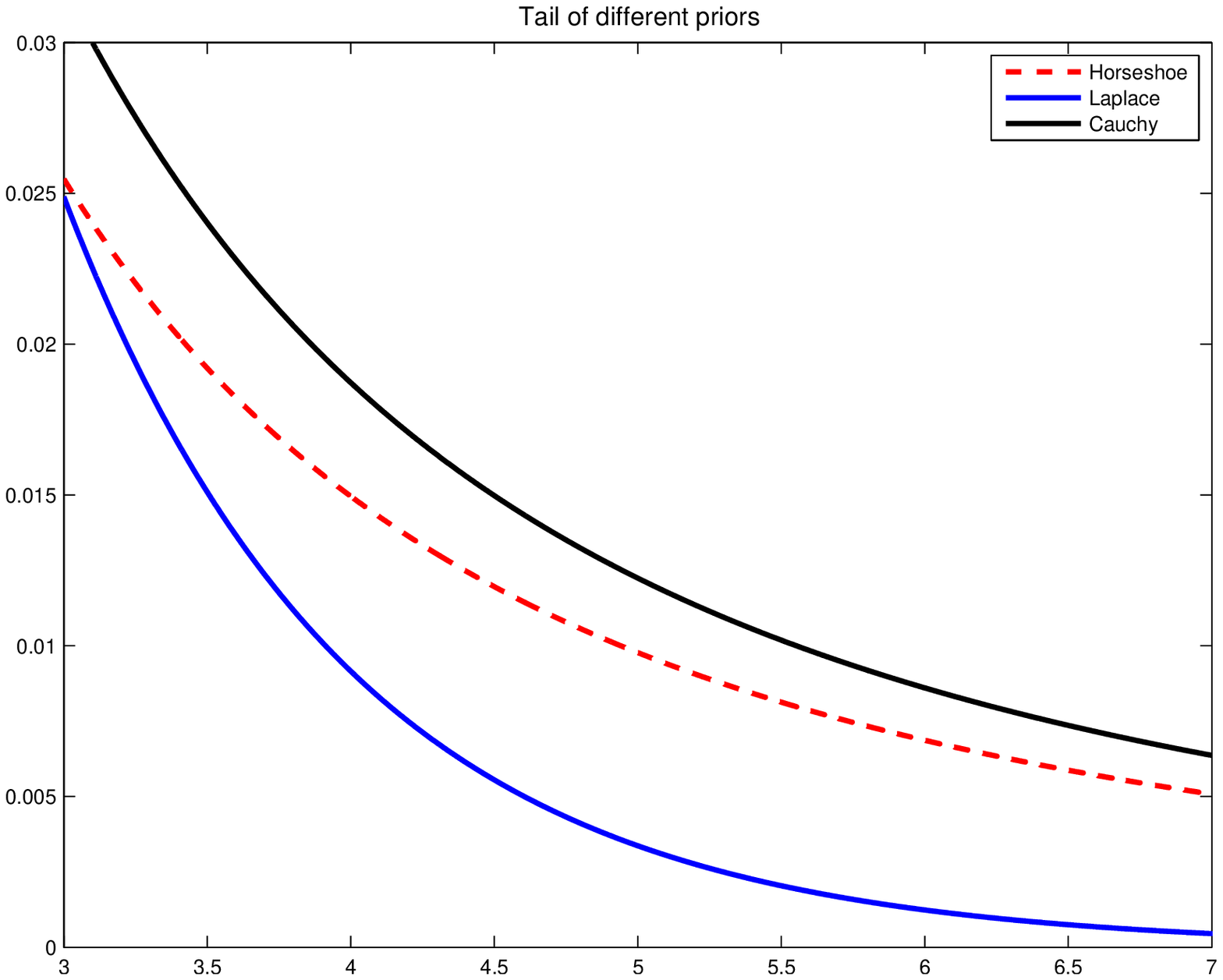}}

\caption[5pt]{A comparison of Horseshoe versus Cauchy and Laplace distribution. (a) compares different distributions in tail, (b) shows Cauchy, Laplace and Horseshoe distribution at origin.} 
\label{Fig.1}   
\end{figure}
    The Horseshoe distribution is a member of GBM family where $ a=\frac{1}{2} $, $ b=\frac{1}{2}$, and can be represented in every model mentioned for GBM. What is more, Horseshoe is also well known as a two-layer hierarchical model in which a half-Cauchy distribution in the second stage is considered on the standard deviation of a normal distribution  ~\cite{9,10,11}.
   \begin{equation*}
        p(\mathbf{w}|\boldsymbol{\gamma})=\prod_{i=1}^{N} \mathcal{N}(w_i|0,{\sigma_i}^2)\ \ p({\sigma})= \textnormal{C}^+(0,1)
     \end{equation*}
    
     In our framework we use the two-layer hierarchical model with a Beta prime distribution in the second stage. Horseshoe has an infinite peak near zero which is the result of having a pole at zero. Also it is almost as heavy-tailed as Cauchy distribution~\cite{10,11}. Figure~\ref{Fig.1} compares the behavior of different distributions in tail and origin.
    \section{Bayesian Inference}\label{section III}
    Bayesian inference analysis is based on the posterior distribution. Since the calculation of $p(\mathbf{w},\boldsymbol{\gamma},\beta|\mathbf{y})$ is computationally intractable, the exact Bayesian inference is not possible, therefore an approximation method should be utilized. In this work, we drive an inference procedure based on the Expectation Maximization method which iteratively estimates the unknown parameters and signal coefficients. Then by inspiration of the Fast-RVM algorithm~\cite{13}, we suggest a reconstruction algorithm detailed in Subsection \ref{3.2.2}.
    \subsection{Explicit Algebraic EM-update}   
 We now formulate the Bayesian inference utilizing the EM method. In this regard we decompose the posterior pdf as :
      \begin{equation}
         p(\mathbf{w},\boldsymbol{\gamma},\beta|\mathbf{y})=p(\mathbf{w}|\mathbf{y},\boldsymbol{\gamma},\beta)p(\boldsymbol{\gamma},\beta|\mathbf{y})\label{(9)}
     \end{equation} 
      The maximization of the posterior distribution on the left hand side of \eqref{(9)} is almost impossible, hence a two-step approach is adopted. In the first step by maximizing $p(\boldsymbol{\gamma},\beta|\mathbf{y})$ the MAP estimation of $\{\boldsymbol{\gamma} , \beta\}$ is computed, while  the second step calculates the conditional MAP estimation of $\mathbf{w}$, with $\{ \boldsymbol{\gamma},\beta\}$ set to values obtained in the first step. 
     Here $\mathbf{w}$ is considered as the hidden variable. The E-step of the EM algorithm aims to compute the conditional expectation
\begin{equation} 
 \mathbf{E}\{ \log{p(\mathbf{y},\mathbf{w},\boldsymbol{\gamma},\beta)} \}
\label{(a)}
\end{equation}
with respect to the conditional distribution $ p(\mathbf{w}|\mathbf{y},\boldsymbol{\gamma},\beta)$ which is a multivariate Gaussian $ p(\mathbf{w}|\boldsymbol{\gamma},\beta,\mathbf{y})=\mathcal{N}(\mathbf{w}|\boldsymbol{\mu},\Sigma) $ with parameters~\cite{4}: 

\begin{equation}
\mu=\Sigma\beta\Phi^Ty \ \ \ \ 
\Sigma={(\beta\Phi^T\Phi+\Lambda)}^{-1}
\label{(11)}
\end{equation} 
where, $\Lambda=\textnormal{diag}({1/\gamma_i})$. The M-step updates the estimation of hyper-parameters $\{\boldsymbol{\gamma} , \beta\}$ by calculating the maximizers of \eqref{(a)}. During the E-step $\{\boldsymbol{\gamma} , \beta\}$ are set to the value already obtained in the M-step. \newline
By using the relation $p(\boldsymbol{\gamma,\beta|\mathbf{y}}) =\frac{p(\boldsymbol{\gamma,\beta,\mathbf{y}})}{p({\mathbf{y}})}\propto p(\mathbf{y},\boldsymbol{\gamma},\beta)$ the estimation of  $\{\boldsymbol{\gamma} , \beta\}$ is computed by maximizing over $p(\mathbf{y},\boldsymbol{\gamma},\beta)$. In order to obtain $p(\mathbf{y},\boldsymbol{\gamma},\beta)$ , we should marginalize out $\mathbf{ w}$, from the joint distribution as follows:
\begin{equation}
p(\mathbf{y},\boldsymbol{\gamma},\beta)=\int p(\mathbf{y},\mathbf{w},\boldsymbol{\gamma},\beta)d\mathbf{w}\nonumber
=\int p(\mathbf{y}|\mathbf{w},\beta)p(\mathbf{w}|\boldsymbol{\gamma}) p(\boldsymbol{\gamma}|a,b) d\mathbf{w}
\end{equation}
Consequently the log-likelihood function is :
\begin{align*}
\mathcal{L}&=-\frac{1}{2}\log|C|-\frac{1}{2}{\mathbf{y}}^T C^{-1}\mathbf{y}\nonumber
+(a-1)\sum\log(\gamma_i) \\
&-(a+b)\sum\log(1+\gamma_i)
-N\log(Beta(a,b))
\label{(14)}
\end{align*}
Where $C$ is defined as $C = (\beta^{-1}\mathbf{I}+\Phi\Lambda^{-1}\Phi^T)$.
 We compute the maximizer of log-likelihood function by taking the derivative of above equation with respect to each hyper-parameter and setting the result equal to zero. Doing so we obtain:
\begin{equation*}
\gamma_i=\frac{\langle w_i^2\rangle+2a-3}{2b+3} \nonumber
+\frac{\sqrt{{(\langle w_i^2\rangle+2a-3)}^2+4(2b+3)\langle w_i^2\rangle}}{2b+3}
\end{equation*}
\begin{equation*}
\beta=\frac{N}{{\Arrowvert y-\Phi\boldsymbol{\mu}\Arrowvert}^2} \label{(16)}
\end{equation*}
 For more details about computing the maximizers of likelihood equation, we refer the reader to section III of~\cite{7}. The mentioned EM algorithm suffers from some drawbacks. For example  the EM method has low convergence rate and a high dimensional matrix inversion is required at each iteration. To address these problems and decrease the computational requirements a fast suboptimal approach was suggested in ~\cite{13}. Within this framework and by considering GBM prior for sparse signals, we suggest a fast iterative sequential algorithm which is summarized in Algorithm\ref{alg1} in the next section.
\subsection{Fast Greedy Update}
\label{3.2}
  The fast algorithm commences with an empty model and sequentially adds the relevant basis. Also as an important feature, in the fast algorithm a basis function can be deleted from the model once it was added~\cite{13}.
  The fast method is relied on efficiently calculating the maximum of the marginal likelihood equation with respect to a single hyper-parameter $ \gamma_i $. Instead of the $ \boldsymbol{\gamma}$ vector, only a single $\gamma_i$ will be updated at each iteration. In this regard, by following the approach in~\cite{13}, dependency of $ \mathcal{L}(\boldsymbol{\gamma})$ on a single hyper-parameter $ \gamma_i $ should be isolated. So we rewrite the likelihood function as:
  \newline
  \begin{align*}
  \mathcal{L}(\boldsymbol{\gamma})&=[-\frac{1}{2}\log|C^{-i}|-\frac{1}{2}\mathbf{y}^T C^{-1}_{-i} \mathbf{y}\nonumber 
  +(a-1)\sum_{i\ne j}\log(\gamma_j)-(a+b)\sum_{i\ne j}\log(1+\gamma_j)]\nonumber \\
  &+[\frac{1}{2}\log{\frac{1}{1+\gamma_i s_i}}+\frac{1}{2}\frac{{q_i}^2\gamma_i}{1+\gamma_i s_i}+(a-1)\log{\gamma_i}\nonumber \nonumber 
  -(a+b)\log(1+\gamma_i)]+c \\
  &=\mathcal{L}({\boldsymbol{\gamma}}_{-i})+{l}({\gamma_i})+c
  \end{align*}
  Where $C_{-i}$ is $C$ when the contribution of $i^{th}$ basis is removed and $ s_i $ and $ q_i $ are defined as :
\begin{equation*}
s_i=\phi_i^TC^{-1}_{-i}\phi_i\ \ \ q_i=\phi_i^TC^{-1}_{-i}\mathbf{y}
\end{equation*}
Since $ C^{-1}_{-i} $ do not depend on $ \gamma_i $, $ s_i $ and $ q_i  $ are independent of $ \gamma_i $ . The terms related to $ \gamma_i $  are now isolated. We are looking for $ \gamma_i $ which maximizes the likelihood equation and at the same time we should consider that the definition region for $ \gamma_i $  is $ \mathbf{R}^+ $. The derivative of likelihood function with respect to $ \gamma_i $ is as follows:
\begin{equation}
\frac{d\mathcal{L}(\boldsymbol{\gamma})}{d\gamma_i} =\frac{dl(\gamma_i)}{d\gamma_i}=-\frac{1}{2}\frac{s_i}{1+s_i\gamma_i}+\frac{1}{2}\frac{q_i^2}{{(1+\gamma_is_i)}^2}\nonumber
+\frac{a-1}{\gamma_i}-\frac{a+b}{1+\gamma_i}
\label{(17)}
\end{equation}
Equating $\frac{dl(\gamma_i)}{d\gamma_i}  $ to zero leads us to a cubic equation with a form of $ h(\gamma_i)=m_1{\gamma_i^3}+m_2{\gamma_i^2}+m_3{\gamma_i}+m_4 $, where
\begin{align*}
& m_1=-(3+2a)s_i^2 \\
 & m_2=(5+4b)s_i+(3-2a)s_i^2-q_i^2\\
& m_3=(5-4a)s_i-q_i^2+2+2b\\
& m_4=2(a-1)
\end{align*}
The roots of $h(\gamma_i)  $ can be determined analytically, but a further analysis is required to realize the maximizer of $ l(\gamma_i) $. Note that, depending on the sign of $ (a-1) $, $ l(\gamma_i) $ diverges either to plus or minus infinity when $ \gamma_i $ tends to zero. But for   $ a\ge 1 $, GBM is not a sparsity-inducing prior so we limit ourselves to $ a<1 $ and continue the inference procedure with this assumption. 
  \subsubsection{The stationary points of $ l(\gamma_i) $}
  \label{3.2.1}
  Since $l(\gamma_i) $ diverges to plus infinity at the origin, it has no global maximum but we can prove that it has at most one local maximum. In the case of existence, we select the local maximizer of $ l({\gamma_i}) $, otherwise  $ {\gamma}_i=0 $ is the maximizer of $ l({\gamma_i})$ and the corresponding basis vector should be pruned from the model.
  \newtheorem{prop}{Proposition}
 \begin{prop}
 if $ a<1 $, $ l(\gamma_i)$ has at most one local maximum.
 \end{prop}
 For $a<1$ since $h(0)<0$ and $ \lim_{\gamma_i\rightarrow\infty}h(\gamma_i)=-\infty $, $ h(\gamma_i) $ has either no positive root or two positive roots. If there exists two positive roots and they are distinct, the greater root is the local maximizer of $ l({\gamma_i})$, otherwise the maximum of $ l({\gamma_i})$ happens at ${\gamma}_i=0 $. 
The above cases can be distinguished through the discriminant of the cubic equation which reveals the nature of the roots. For a cubic equation , the discriminant is defined as $ \Delta=18 m_1m_2m_3m_4-4m_2^3m_4+m_2^2m_3^2-4m_1m_3^3-27 m_1^2m_4^2 $. For $ \Delta >0 $ the equation has three distinct real roots, for $ \Delta=0 $ a multiple root when all roots are real and if $  \Delta <0 $ the equation has one real root and two non-real complex conjugate roots. Provided that the discriminant of  $ h(\gamma_i ) $ is positive, it has either three negative distinct roots or two positive and one negative root. The two positive roots happens in two scenarios:
\begin{enumerate}[I]
\item $ \Delta>0$ and $ m_3>0 $, in this case the stationary points of $ h(\gamma_i ) $ lies in the different sides of $ y- $axis, so the three roots include positive roots.\newline
\item $\Delta>0, m_3<0 $ and $ m_2<0 $,	in this case the stationary points of $ h(\gamma_i) $ have the same sign but since $ m_2<0 $, the sum of them is positive. Hence we conclude that $ h(\gamma_i ) $ has two positive roots.\newline
\end{enumerate}
\subsubsection{Fast Algorithm}
\label{3.2.2}
Based on the inference procedure detailed in Subsection\ref{3.2}, we suggest a greedy algorithm which is summarized in Algorithm\ref{alg1}. The free parameters of GBM, $  a $ and $ b $, are set manually at step 1 of the algorithm. In order to reach a higher performance, the free parameters should be tuned by considering the degree of signal sparsity and non-zero coefficients amplitude. At step 5 of the algorithm a basis function should be selected from the basis functions included in or excluded from the model for updating. The selecting procedure could be done randomly or on an order. Following the approach in~\cite{13}, we calculate the change in marginal likelihood for each potential candidate. The $ \gamma_i $, which results in the greatest change will be chosen for updating. At each iteration along with $ \gamma_i $, the value of $  s_i $ , $ q_i $ , $ \mu $ and $ \Sigma$ should also be updated. For the sake of computational efficiency, we follow the method in RVM formulation for updating  $  s_i $ , $ q_i $ , $ \mu $ and $ \Sigma$ in the steps 12 and 13 (see Appendix A of~\cite{13} for more details), but due to using new priors the likelihood equation and the updating procedure for $ \gamma_i $ is quite different from the similar algorithms in~\cite{7,13}.
\begin{algorithm}[h]
\caption{Fast-GBM}
\label{algo1}
\begin{algorithmic}[1]

\State {$\mathbf{{Inputs}}:\Phi,\mathbf{y},a<1,b$}
\State {$ \mathbf{{Outputs}}:\mathbf{w},\boldsymbol{\gamma},\Sigma $}
\State $ \mathbf{Initiate}: \boldsymbol{\gamma}=0$
\While {convergence criterion not met}
\State{Choose a $ \gamma_i $}
\If {$\Delta>0 \And m_3>0\  | \ \Delta>0 \And m_3<0 \And m_2>0$} \If {$ \gamma_i=0 $} {Choose the greater root as the new update for $ \gamma_i $ and add $ \phi_i $ to the model}\ElsIf {$ \gamma_i>0 $} {Choose the greater root as new update for $ \gamma_i $}
\EndIf
\Else \ {Prune $ i^{th} $ basis from the model (set $ \gamma_i=0 $)}
\EndIf\ {stop criterion}
\State{$ \mathbf{Upate}\  \Sigma,\boldsymbol{\mu} $}
\State{$ \mathbf{Upate}\ s_i,q_i $}

\EndWhile

\end{algorithmic}
\label{alg1}
\end{algorithm}
\newtheorem{thm}{Remark}
\begin{thm}
As mentioned in section~\ref{section II},  Horseshoe is GBM distribution with specified  hyper-parameters ($ a=b=\frac{1}{2} $). Since $ a=\frac{1}{2} $ and less than 1, the inference procedure for Horseshoe is exactly
the same as GBM; therefore we should just simply substitute $a=b=\frac{1}{2}$ in the likelihood equation.
\end{thm}

\section{Experimental Validation}\label{section IV}
In this section, we present simulation results to analyze the performance of the proposed algorithms in Subsection \ref{3.2.2}. We compare the performance of our algorithms with the other state-of-art sparse estimators in the Bayesian framework, Fast-BCS\footnote[1]{\url{http://people.ee.duke.edu/~lcarin/BCS.html}} and Fast-Laplace\footnote[2]{\url{http://ivpl.eecs.northwestern.edu/}}. We consider a signal of dimension $ N $ where T coefficients at random location are drawn from uniform distribution and the other $ N-T $ coefficients are set to zero. The measurement matrix $ \Phi $ is chosen to be a uniform spherical ensemble. Through out every set of the experiments, we fix $ N=1000 $ and $ SNR=40db $. We also repeat each experiment 100 times and report the  average result. As an error measure we take the relative MSE ${ \Arrowvert\hat{\mathbf{w}}-\mathbf{w}\Arrowvert_2^2}/{\Arrowvert\mathbf{w}\Arrowvert_2^2} $ where $ \hat{\mathbf{w}} $ and $ \mathbf{w} $ are the estimated and original signal vector respectively. We present analyses with respect to three parameters
 \paragraph{1. Performance with Respect to Different Levels of Signal Amplitude:} In the first set of our experiments, we aim to  evaluate the ability of the algorithms in reconstructing signals with different amplitudes. In this regard, we fix $ M=300 $ and $ T=30 $, while the non-zero coefficients amplitude varies from $ 0 $ to $ A $. $ A $  is also varied from 10 to 100 in the steps of 10. We report the reconstruction error, estimated number of non-zero coefficients $ \hat{T} $ and number of the required iterations for algorithm convergence. The results are demonstrated in Fig.\ref{amplitude}. As it has been shown in Fig.3(c) for the signals with the amplitude of less than approximately 35, Fast-Laplace and Fast-BCS show better performance in terms of reconstruction error, while for $ A>35 $ our proposed algorithms lead to a lower reconstruction error. This is due to the heavy-tailed properties of GBM and Horseshoe which model coefficients with larger amplitudes more efficiently. Fig.3(b) demonstrates that Fast-Laplace and Fast-BCS fail in estimating the number of non-zero coefficients when the amplitude of spikes is larger than 20, while our proposed algorithms give a reasonable estimation of $ T $ for all amounts of the signal amplitude in this range. It is obvious from Fig.3(a) that Fast-GBM and Fast-Horseshoe have faster convergence rate. 
 \begin{figure}
                  \centering
              \subfigure[]{\includegraphics[width=35mm,height=28mm]{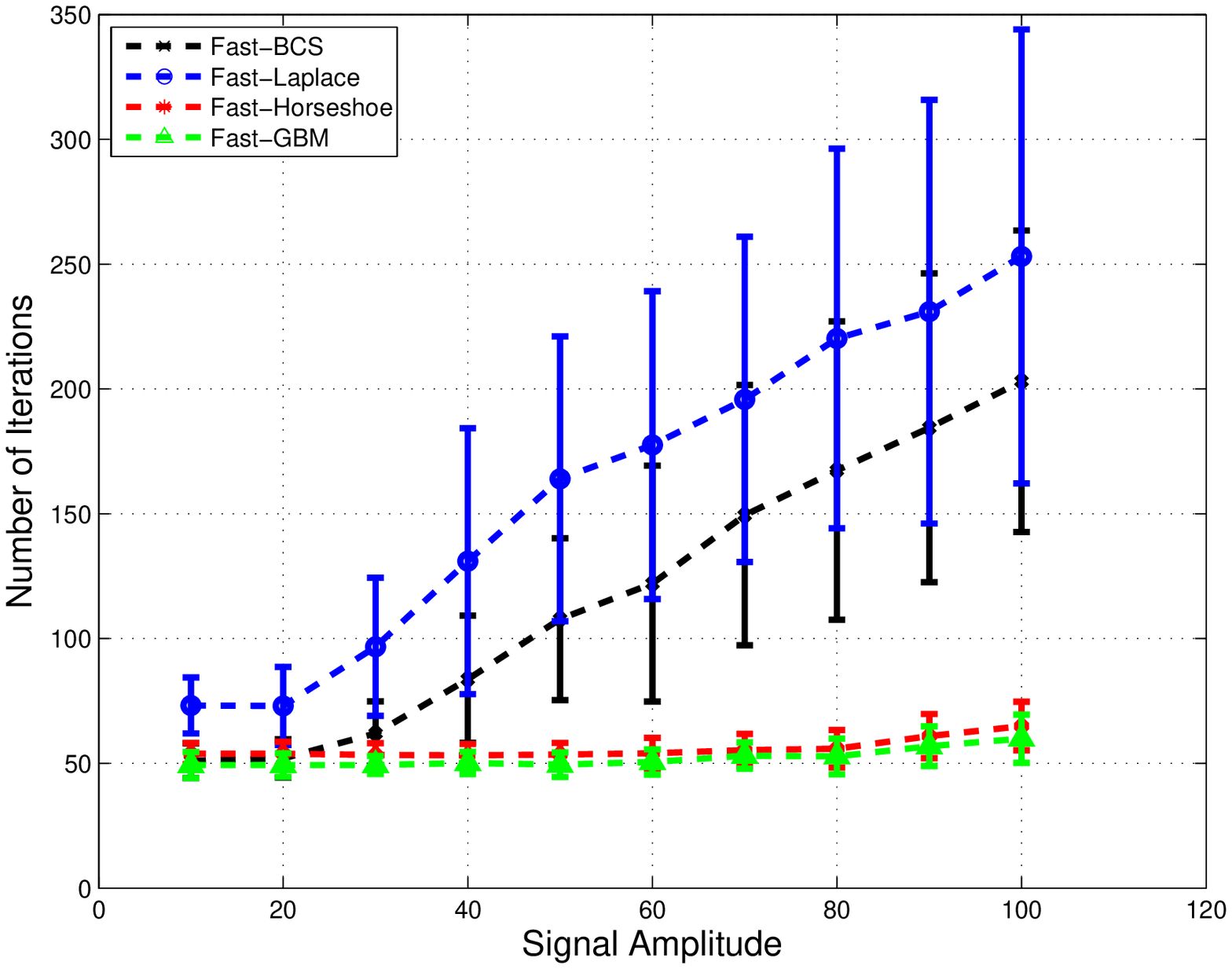}}\label{N-A}
                \subfigure[]{\includegraphics[width=35mm,height=28mm]{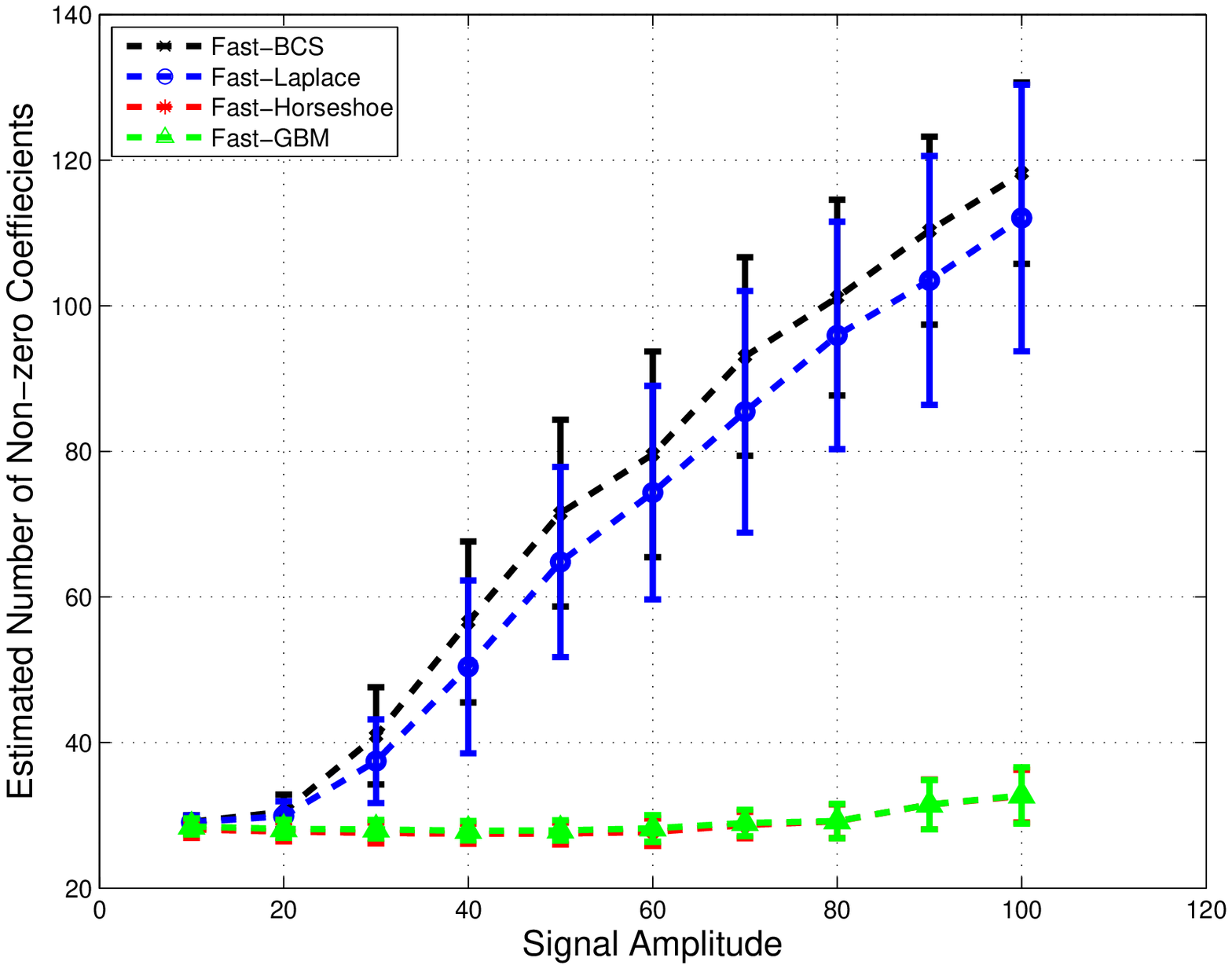}}\label{T-A}
                \subfigure[]{\includegraphics[width=35mm,height=28mm]{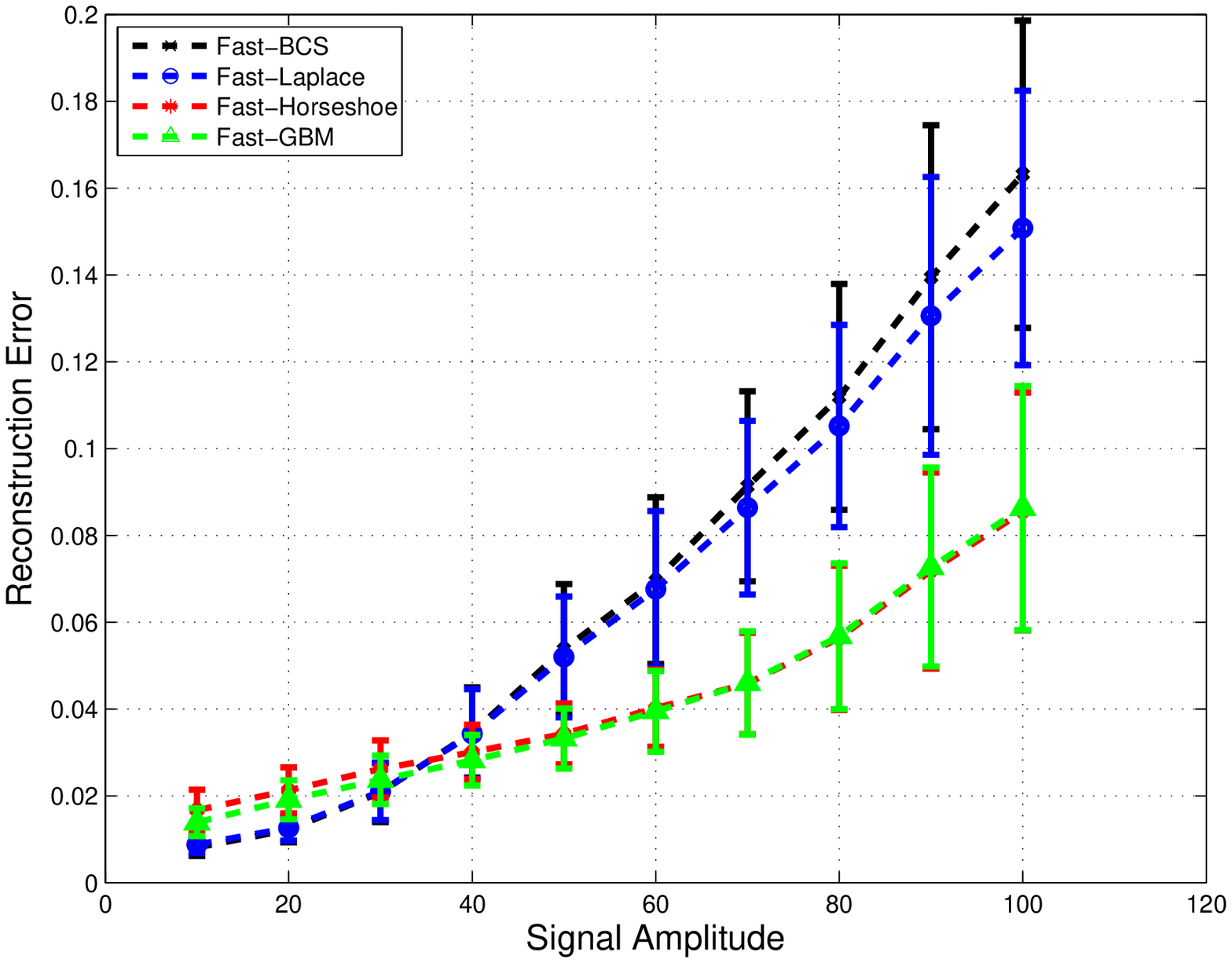}}\label{SA}
                \caption[5pt]{Signal coefficients amplitude (A) versus algorithms performance. (a) shows A versus number of iterations for algorithm convergence, (b)A versus estimated number of non-zero coefficients, (C) A versus mean square  reconstruction error.}
                \label{amplitude}    
                \end{figure} 
   
 \paragraph{2. Performance with Respect To Different Number of Measurements:} In the second set of our experiments, we study the performance of the algorithms as a function of the number of measurements  $ M $. In these investigations, we fix $ T=30 $ and vary the number of measurements $ M $ from $ 240 $ to $350  $ in steps of $ 10 $ and measure the reconstruction error, the estimated number of non-zero coefficients and the required numbers of iterations for algorithm convergence. The results are shown in Fig.\ref{M}. Here the amplitude of random spikes varies between 0 and 35 which corresponds to where all of the curves meet in Fig.3(c). It means that for this amplitude all of the algorithms show similar performance in terms of reconstruction error. The simulation results reveal that  Fast-GBM outperforms the others in terms of reconstruction error except for $ M>330 $, that Fast-Laplace and Fast-BCS show better performance. The results in Fig.4(a) demonstrates that the convergence rate of Fast-GBM and Fast-Horseshoe is roughly independent of the number of measurements and clearly our proposed algorithms require fewer iterations. Fig.4(b) shows that the estimated number of non-zero coefficients returned by Fast-GBM and Fast-Horseshoe seem to be almost constant and considerably lower compared to the obtained results by Fast-BCS and Fast-Laplace.
 
 \begin{figure}
      \centering
      
    \subfigure[]{\includegraphics[width=35mm,height=28mm]{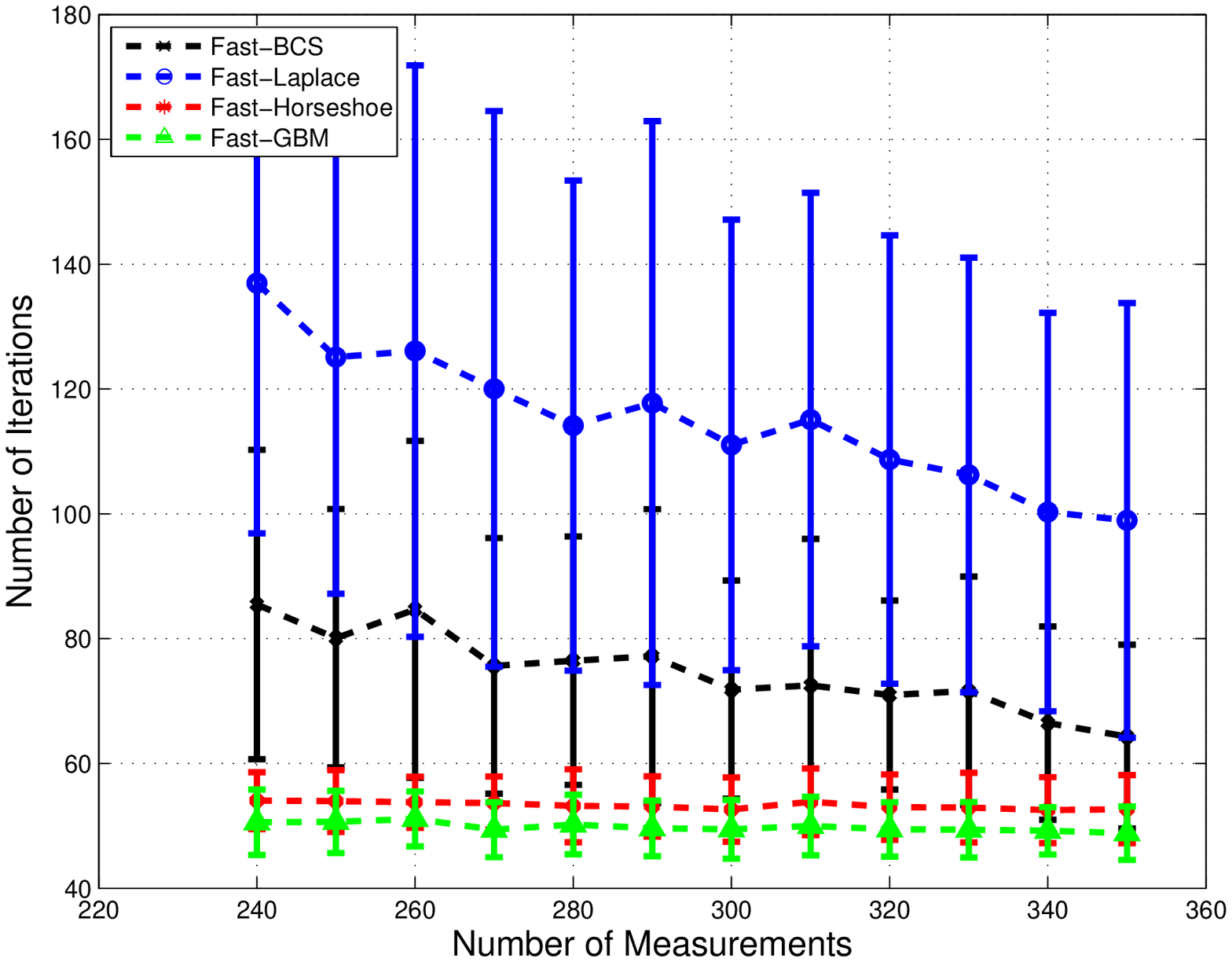}}
    \subfigure[]{\includegraphics[width=35mm,height=28mm]{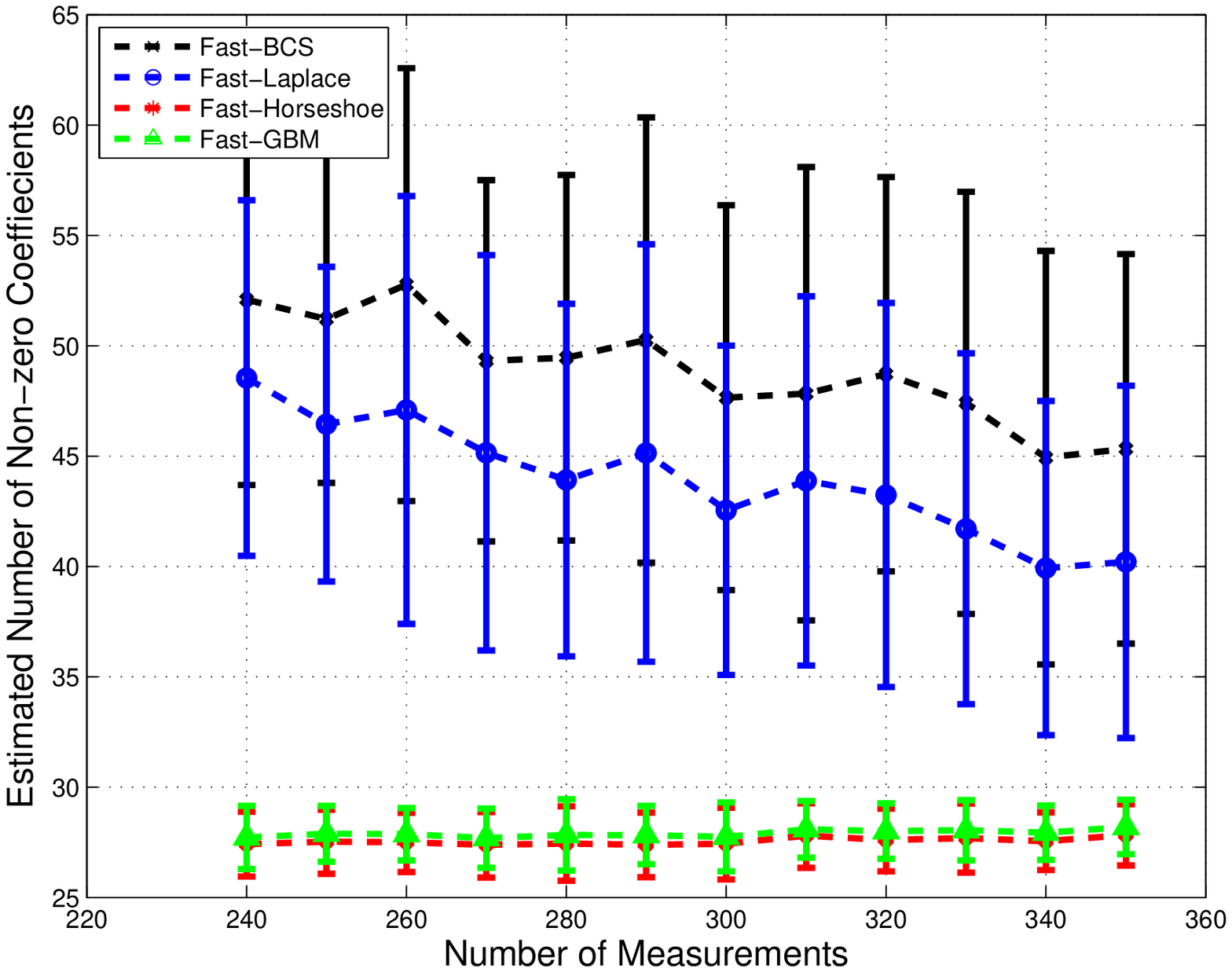}}
    \subfigure[]{\includegraphics[width=35mm,height=28mm]{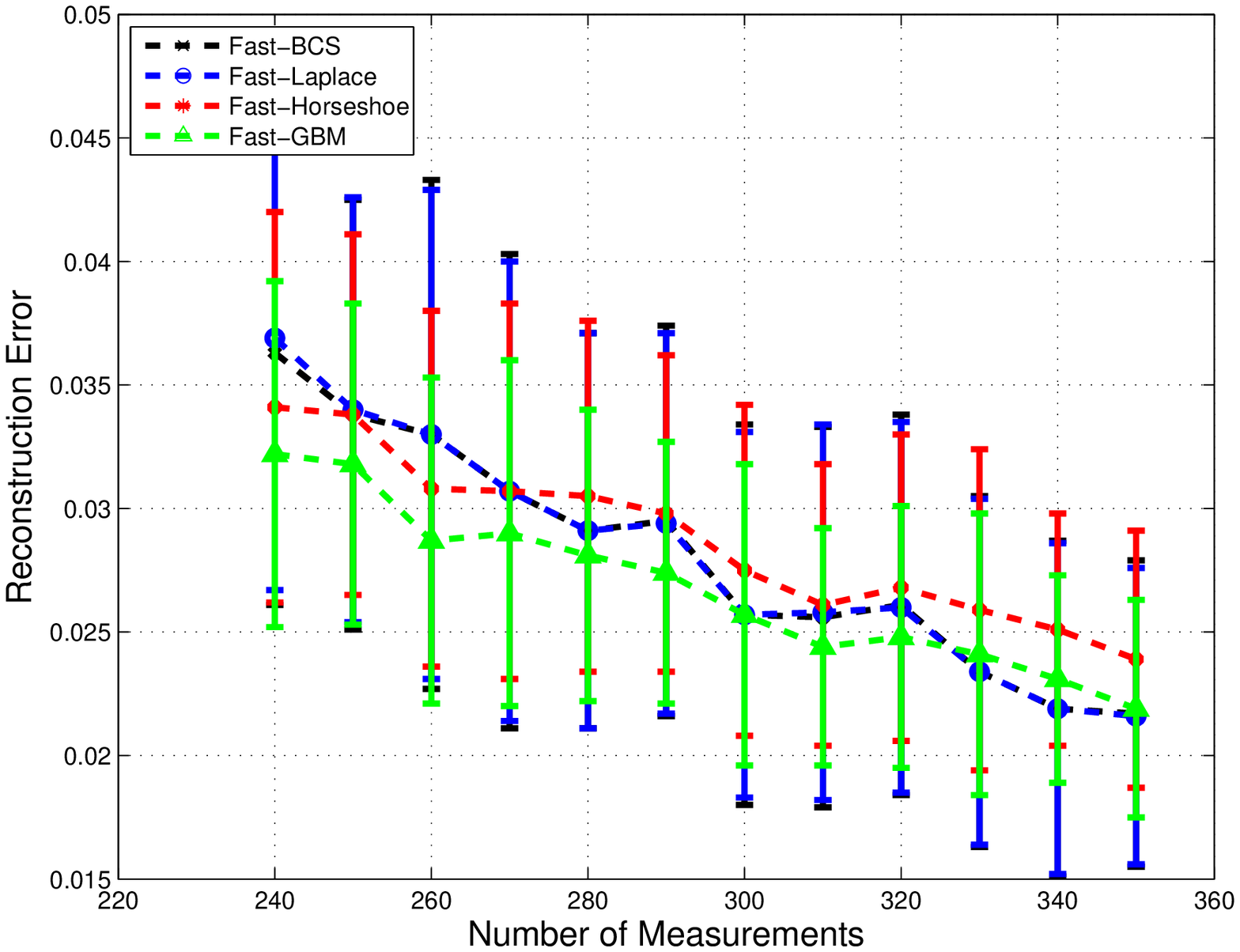}}
    \caption[5pt]{Number of measurements (M) versus algorithms performance. (a) shows M versus number of iterations for algorithm convergence, (b)M versus estimated number of non-zero coefficients, (C) M versus mean square reconstruction error.The amplitude of spikes is between 0 and 35.}  
    \label{M}
  \end{figure}
  \begin{figure}[t]
                \centering
                
              \subfigure[]{\includegraphics[width=35mm,height=28mm]{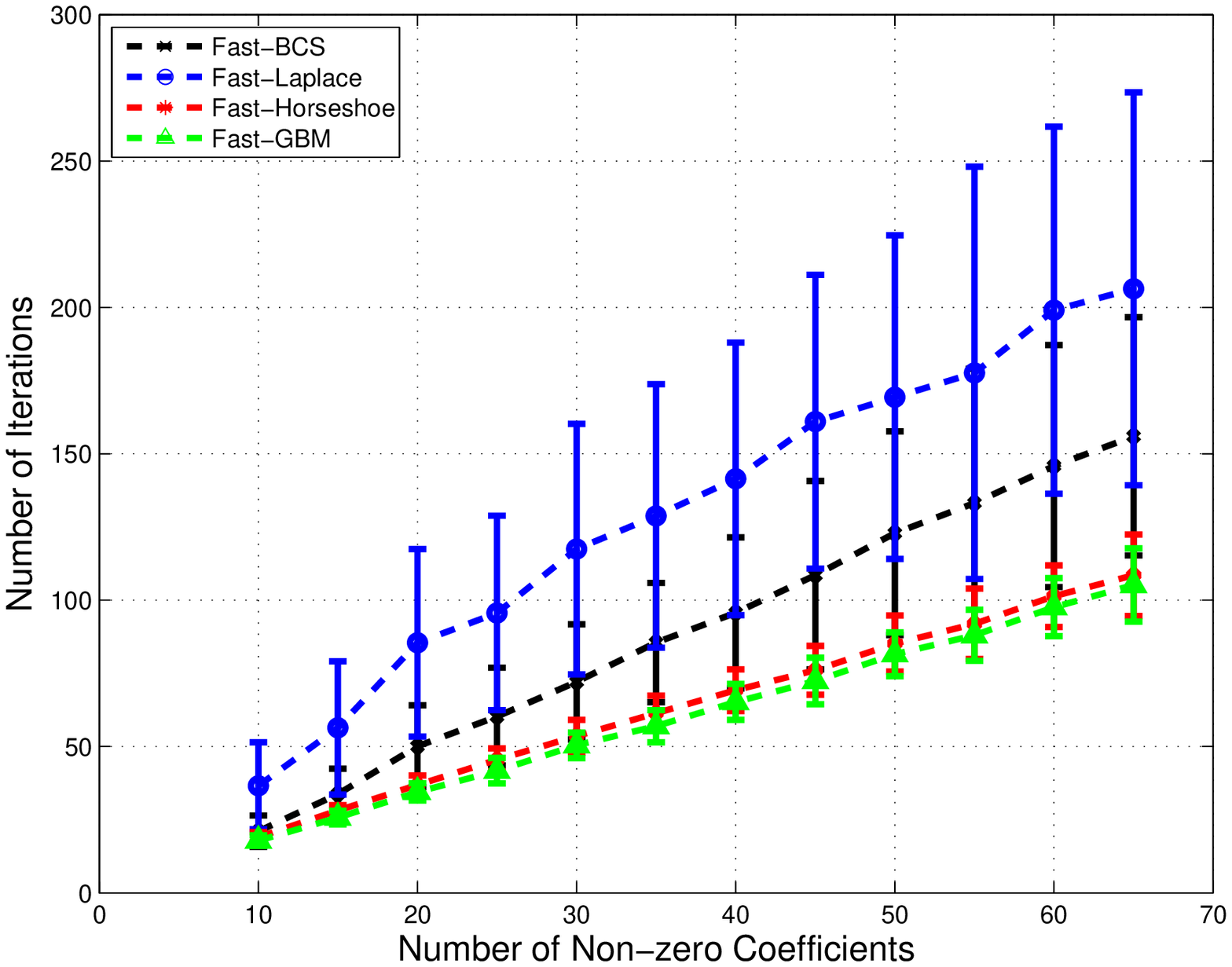}}
              \subfigure[]{\includegraphics[width=35mm,height=28mm]{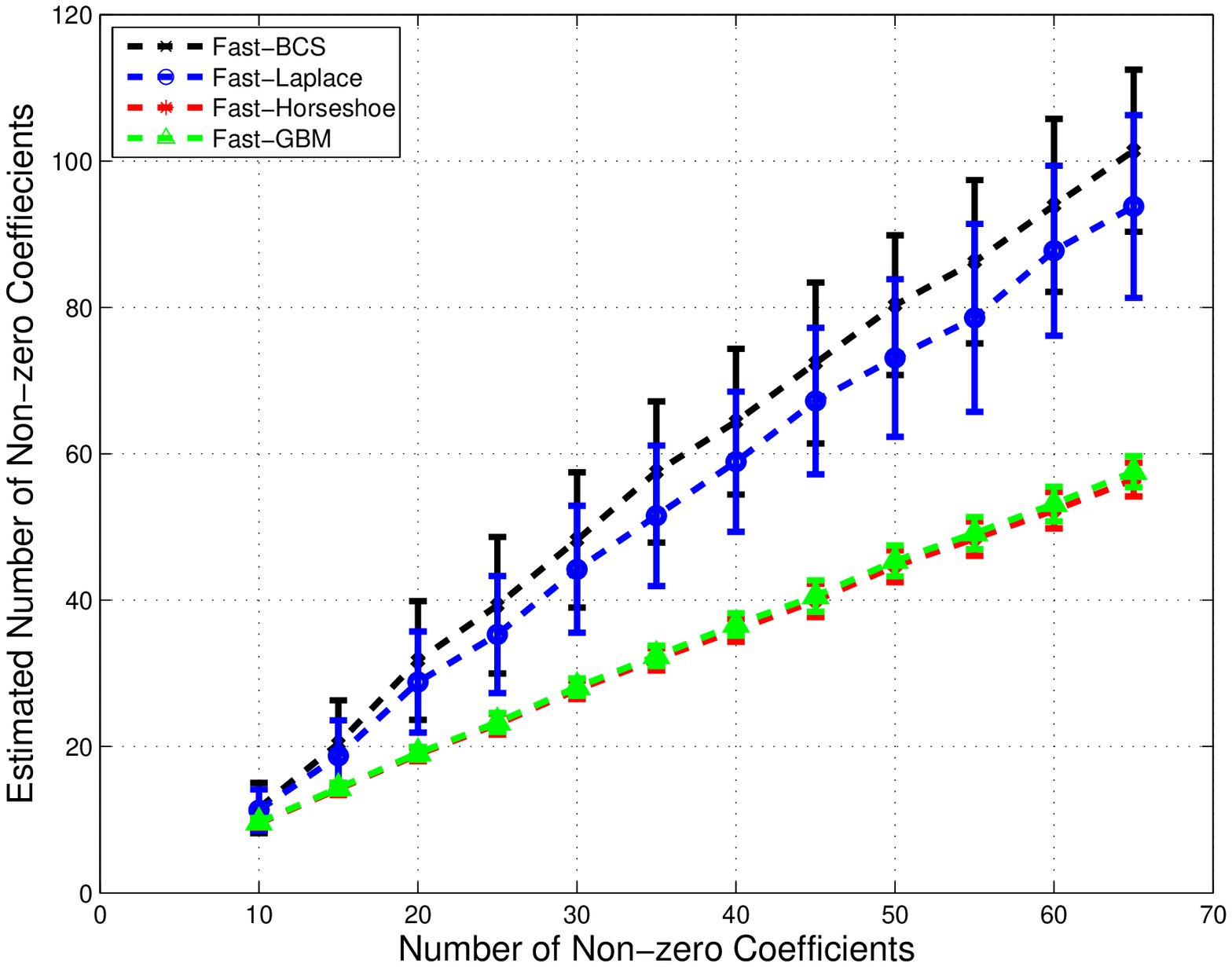}}
              \subfigure[]{\includegraphics[width=35mm,height=28mm]{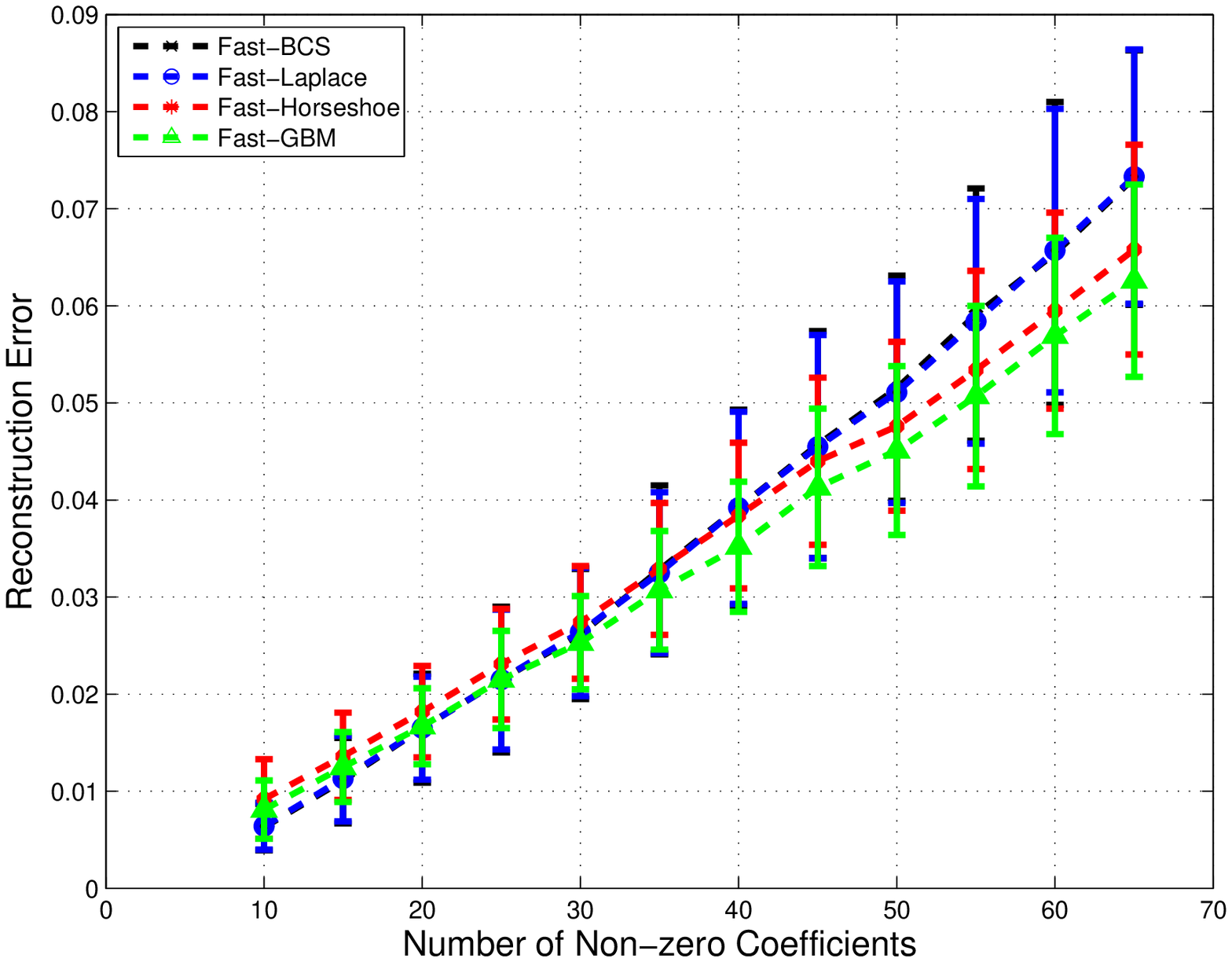}}
              \caption{Number of non-zero coefficients (T)versus algorithms performance.(a)shows T versus number of iterations for convergence, (b) T versus estimated number of non-zero coefficients,(C) T versus mean square reconstruction error. The amplitude of spikes is between 0 and 35.}    
              \label{T}
              \end{figure}
    \paragraph{3. Performance with Respect to Different Levels of Signal Sparsity:} The third set of experiments is designed to analyze the performance of the algorithms in handling different levels of sparsity. In this regard we vary $ T $ from 10 to 65 in the steps of 5, while we fix $ M=300 $ and the amplitudes of spikes are between 0 and 35 randomly. The corresponding performance results are depicted in Fig.~\ref{T}. It can be seen from Fig.5(a) and Fig.5(c) that, while all of the algorithms show similar performance in terms of reconstruction accuracy, our algorithms have faster convergence rate. The results in Fig.5(b) show that Fast-GBM and Fast-Horseshoe considerably outperform the other algorithms in providing an accurate and sparse estimation of $ T $. This is due to the interesting behaviour of the Horseshoe and GBM distributions at origin which enforces sparsity more heavily. 
    \newline
    In summary, the simulation results suggest that the performance of our proposed framework is competitive when it is applied to the sparse signals with larger amplitude. However the Fast-GBM and Fast-Horseshoe algorithms provide improved performance in terms of convergence rate and accurate estimation of the number of non-zero coefficients in all of the cases.
    \section{Conclusion}\label{V}
       In this paper we considered the compressive sensing problem in Bayesian framework. We reviewed the existing sparsity-inducing priors in the literature. As new priors, we suggested the Horseshoe and Generalized Beta Mixture as  potential candidates for modeling sparse signals in Bayesian structure and studied different hierarchical models for these priors. By considering the GBM and Horseshoe as a prior distribution and using the Expectation Maximization inference method, we provided a solution to CS problem and proposed two reconstruction algorithms resulting from this formulation. The simulation results show our proposed algorithms outperform the other state-of-art algorithms in modeling and estimating signals with different amplitude and sparsity level. Our algorithms also have faster convergence rate and provide sparser solutions.

 \end{document}